\documentclass[aps,prb,twocolumn,groupedaddress]{revtex4-1}

\usepackage{graphicx}
\usepackage{hyperref}
\usepackage{multirow}
\usepackage{array}
\usepackage{amsmath}

\begin{document}

\newcommand{\FeAs}{Ba$_{0.72}$K$_{0.28}$Fe$_2$As$_2$}
\newcommand{\FeCoAs}{Ba(Fe$_{0.95}$Co$_{0.05}$)$_2$As$_2$}

\title{Precision Microwave Electrodynamic Measurements of K- and Co-doped BaFe$_2$As$_2$}

\author{J. S. Bobowski}
\author{J. C. Baglo}
\author{James Day}
\author{P. Dosanjh}
\author{Rinat Ofer}
\author{B. J. Ramshaw}
\author{Ruixing Liang}
\author{D. A. Bonn}
\author{W. N. Hardy}
\email[]{hardy@phas.ubc.ca}
\affiliation{Department of Physics and Astronomy, University of British Columbia, Vancouver, British Columbia, Canada V6T 1Z1}

\author{Huiqian Luo}
\author{Zhao-Sheng Wang}
\author{Lei Fang}
\author{Hai-Hu Wen}
\affiliation{National Laboratory for Superconductivity, Institute of Physics and National Laboratory for Condensed Matter Physics, P.O. Box 603 Beijing, 100190, P. R. China}

\date{\today}

\begin{abstract}

We have studied the microwave electrodynamics of single crystal iron-based superconductors Ba$_{0.72}$K$_{0.28}$Fe$_2$As$_2$ (hole-doped, $T_\mathrm{c}$~$\approx~$30~K) and Ba(Fe$_{0.95}$Co$_{0.05}$)$_2$As$_2$ (electron-doped, $T_\mathrm{c}$~$\approx$20~K), by cavity perturbation and broadband spectroscopy. SQUID magnetometry was used to confirm the quality and homogeneity of the samples under study. Through cavity perturbation techniques, the temperature dependence of the in-plane London penetration depth $\Delta\lambda(T)$, and therefore the superfluid phase stiffness $\lambda^2(0)$/$\lambda^2(T)$ was measured. Down to 0.4~K, the data do not show the exponential saturation at low temperatures expected from a singly-, fully-gapped superconductor. Rather, both the electron- and the hole-doped systems seem to be best described by a power law behavior, with $\lambda^2(0)$/$\lambda^2(T)$~$\sim$~$T^n$ and \emph{n}~$\approx$~2.5. In the three samples we studied, a weak feature near the sensitivity limit of our measurements appears near $T/T_\mathrm{c}$~=~0.04, hinting at a corresponding low energy feature in the superconducting density of states. The data can also be relatively well-described by a simple two-gap $s$-wave model of the order parameter, but this yields parameters which seem unrealistic and dependent on the fit range. Broadband surface resistance measurements reveal a sample dependent residual loss whose origin is unclear. The data from the \FeAs~samples can be made to scale as~$\omega^2$ if the extrinsic loss is treated as an additive component, indicating large scattering rates. Finally, the temperature dependence of the surface resistance at 13~GHz obeys a power law very similar to those observed for $\Delta\lambda(T)$.

\end{abstract}


\maketitle

\section{Introduction}

Tremendous interest was generated when the fluorine-doped layered compound LaFeAsO$_{1-x}$F$_x$ was reported to superconduct at 26~K.~\cite{Kamihara08-3296} In remarkably short time, the critical temperature of this compound was increased via pressure or chemical substitution to above 55~K, significantly higher than the highest $T_\mathrm{c}$ reported in any superconductor outside the cuprates (cf., MgB$_2$~\cite{Nagamatsu01-63}). Superconductivity has since been found in Ba$_{1-x}$K$_x$Fe$_2$As$_2$ with $T_{\mathrm{c,max}}$~=~38~K,~\cite{Rotter08-020503} and in Ba(Fe$_{1-x}$Co$_x)_2$As$_2$ with $T_{\mathrm{c,max}}$~=~23~K.~\cite{Ni08-214515} These so-called 122-compounds are particularly important since, unlike the cuprates or the 1111 iron pnictides, they are not oxides, eliminating the potentially problematic role of oxygen stoichiometry. Moreover, large single crystals with a variety of different cation dopings in the 122-pnictides have now been synthesized, which is essential for applying a wide range of measurements of their physical properties. The need to understand the pairing mechanism and the origin of the high $T_\mathrm{c}$ in any new superconductor drives a need to determine the symmetry of the order parameter, however this usually requires several different measurements to arrive at a consensus. Such measurements need to be performed on single-phase samples with well-characterized stoichiometry and sharp transitions, which imply good homogeneity; otherwise, it is challenging to compare one measurement to the next. To go even further, cation and anion disorder should also be minimized and structural information (e.g., x-ray rocking curves) should be obtained.

Historically, one powerful class of measurements that can be thought of as topological include the flux quantization measurements~\cite{Keene87-855} that show superconductors are a condensate of pairs, the phase coherence measurements in bimetallic dc SQUIDs and tunnel junctions made from single crystals of YBa$_2$Cu$_3$O$_{6.8}$ and thin films of the conventional $s$-wave superconductor Pb,~\cite{Wollman93-2134} and the observation of half flux quanta in geometrically frustrated junctions,~\cite{Tsuei94-593} which were decisive in proving the extra broken symmetry in the $d_{x^2-y^2}$ state of the cuprates. Another group of measurements directly probes the superconducting pairing gap via spectroscopic means. This was famously the case in conventional $s$-wave superconductivity in which measurements such as tunneling,~\cite{Giaver60-147} infrared~\cite{Ginsberg60-990,Richards60-575} and microwave spectroscopy~\cite{Turneaure68-4417} showed a well-defined gap with a sharp threshold energy and essentially no states below the energy gap at low temperatures and in the absence of pair-breaking magnetic impurities. Such measurements were more ambiguous in the cuprates because the presence of nodes in the $d_{x^2-y^2}$ pairing state gave a characteristic energy gap scale, but without the very sharp threshold and with many states available down to low energies. Angle-resolved photoemission~\cite{Shen93-1553,Ding96-R9678} helped resolve this by showing the gap variation as a function of momentum around the Fermi surface.

Other measurements rely on inferring the presence of a gapped spectrum of excitations in the system by observing the temperature dependence of a wide range of features, including thermodynamic, transport, and electrodynamic properties. We will concentrate on electrodynamic characteristics here but will start with an introductory comment on the difficulties of making inferences from temperature dependencies in these properties. In conventional $s$-wave superconductivity, the presence of exponentially activated behavior in many properties at low temperatures signals the presence of a non-zero minimum energy gap and hence no nodes. Such measurements~\cite{Corak56-656,Corak56-662} even predate BCS theory, but it has always been difficult to do this decisively since it takes high resolution data, preferably over a few decades of the low temperature exponential behavior, to be convincing. This has, for instance, been achieved in high resolution measurements of the temperature dependent microwave loss of high Q resonant cavities made of Pb.~\cite{Turneaure,Halbritter70-466}

In the case of non $s$-wave states with nodes, the temperature dependencies tend towards various power laws. Here there is a significant challenge in identifying the particular state, or even in being sure that it is not really exponentially activated. One case has proven relatively easy: the line nodes of a $d_{x^2-y^2}$ pairing state on a cylindrical Fermi surface gave rise to an unambiguous linear temperature dependence in the London penetration depth.~\cite{Hardy93-3999} Unfortunately, disorder quickly changes this to a quadratic temperature dependence,~\cite{Hirschfeld93-4219} so that in materials with pair breaking defects, especially cation doping, these techniques place high demands on resolution and careful comparison of power laws versus exponentials, plus considerable systematic work on sample dependence and multiple materials within a family. An example of this can be seen in the long effort to understand the penetration depth in the electron-doped cuprates Pr$_{2-x}$Ce$_x$CuO$_{4-\delta}$ and Nd$_{2-x}$Ce$_x$CuO$_{4-\delta}$.~\cite{Kokales00-3696}

Universal consensus regarding the gap symmetry in the iron-based superconductors does not currently exist, but the field is working hard towards remedying this situation. For example, it is known that the antiferromagnetic ground state in the BaFe$_2$As$_2$ parent compound is suppressed through cation substitution, thus allowing superconductivity to emerge;~\cite{Ni08-214515,Rotter08-107006} being near to a magnetic state might mean that magnetic fluctuations are important for pairing and this could be reflected in the symmetry of the gap. What, then, is the pairing symmetry?

Band structure calculations~\cite{Barzykin08-131,Cvetkovic09-37002} and angle-resolved photoemission spectroscopy experiments~\cite{Liu08-177005,Evtushinsky09-054517} demonstrate that multiple bands cross the Fermi surface, making multi-band superconductivity plausible. For the hole-doped Ba$_{1-x}$K$_x$Fe$_2$As$_2$ compound, ARPES~\cite{Ding08-47001,Zhao08-4402,Nakayama09-67002} has found at least two different superconducting nodeless gaps in the $ab$-plane. These results are further supported by directional point-contact Andreev-reflection spectroscopy~\cite{Szabo09-012503} and microwave surface impedance~\cite{Hashimoto09-207001} data which suggest fully-, and perhaps multiply-, gapped superconductivity. However, the possibility for a nodal gap has not been completely ruled out: measurements of reversible magnetization~\cite{Salem09-014518} and thermal Hall conductivity~\cite{Checkelsky} both yield results consistent with nodes in the gap; $^{75}$As nuclear magnetic resonance measurements~\cite{Fukazawa09-033704} revealed the spin-lattice relaxation rate 1/$T_1$ to vary close to $T^3$; and, muon spin-relaxation measurements~\cite{Goko09-024508} exhibit a nearly linear variation in temperature of the superfluid density at low temperatures. For the electron-doped Ba(Fe$_{1-x}$Co$_x)_2$As$_2$ compound, heat transport measurements~\cite{Dong10-094520} suggest a nodeless superconducting gap in the $ab$-plane. However, tunnel diode resonator techniques~\cite{Gordon09-100506(R)} have revealed that the penetration depth as a function of temperature exhibits a robust power law (instead of exponential) behavior, with $\Delta\lambda(T)$~$\propto$~$T^n$ and \emph{n} being between 2 and 2.5, depending on the doping level. The question of pairing symmetry remains open.

\section{Materials and Methods}

\begin{figure}[htpb]
\begin{center}
\includegraphics[bb= 65 25 665 535, width=\columnwidth]{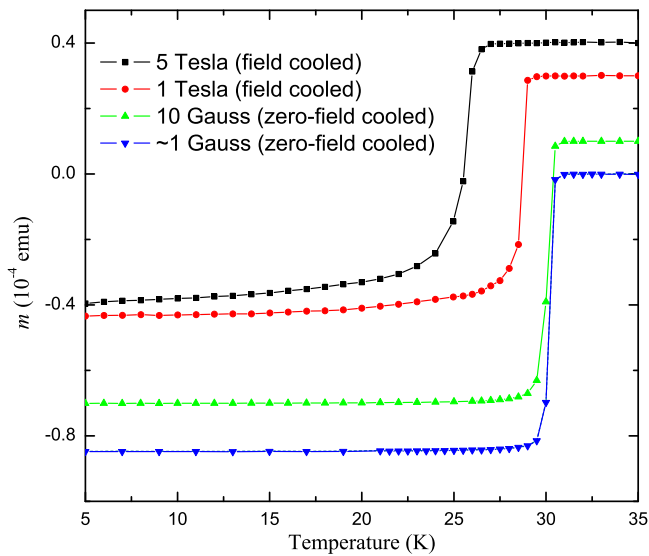}
\caption{(Color online) The magnetic moment $m$ of the \FeAs~sample A, as measured by a (Quantum Design MPMS) SQUID magnetometer. The data have been multiplicatively scaled to fit on the same plot. Even at high applied magnetic fields, the width of the superconducting transition remains sharp.} \label{fig:SQUIDdata}
\end{center}
\end{figure}

In this paper, we report on measurements of the temperature dependence of the London penetration depth and surface resistance in the hole-doped \FeAs~and the electron-doped \FeCoAs~122-compounds. We have measured three high-quality single crystals, grown by the Wen group at the National Lab for Superconductivity in Beijing, using an FeAs self-flux method:~\cite{Luo08-125014} two of \FeAs~and one of \FeCoAs, with their dimensions listed in Table~\ref{tab:1}. The K-dopant occupies out-of-plane interstitial sites in the crystal lattice, whereas Co substitutes for Fe in the Fe$_2$As$_2$ plane. Our microwave techniques are optimized for 1~mm$^2$ platelets and the samples used in these measurements were carefully selected to be the best single crystals available. The sample surfaces are known to degrade from prolonged exposure to ambient atmosphere. To limit surface degradation, we have only measured samples with cleaved $ab$-surfaces and between measurements the samples were stored in a vacuum desiccator. Furthermore, these samples have little secondary impurity phase (less than 10\%) as checked by specific heat.~\cite{Mu09-174501} Sample quality and homogeneity were confirmed via the width of the superconducting transition as a function of applied field, as shown in Fig.~\ref{fig:SQUIDdata}. The magnetic moment $m$ of sample A, for example, was measured in dc magnetic fields applied parallel to the $\hat c$-axis of the crystal. In low fields ($<$~10~G), $T_\mathrm{c}$~=~30.1~K and $\Delta T_\mathrm{c}$~$<$~0.5~K. At 5~Tesla, $T_\mathrm{c}$ is suppressed by 15\% to 25.6~K, but the transition width remains narrow ($\Delta T_\mathrm{c}$~$<$~1.5~K): a clear signature of a homogeneously doped sample.

\begin{table}[ht]
\caption{Transition temperatures and sample dimensions of the iron-arsenide crystals studied in this paper.}\label{tab:1}
 \begin{center}
  \begin{tabular}{|c|c|c|c|c|}
    \hline
    Sample & ~$T_\mathrm{c}$~ & ~$ab$-surface~ & ~$\hat c$-axis~ & ~aspect~\\
    ~ & ~(K)~ & ~(mm$^2$)~ & ~($\mu$m)~ & ~ratio~\\
    \hline
    (Ba,K)Fe$_2$As$_2$ (A) & ~29.5~ & ~0.928~ & ~50~ & ~20~ \\
    (Ba,K)Fe$_2$As$_2$ (B) & ~28~ & ~0.508~ & ~10~ & ~50~ \\
    Ba(Fe,Co)$_2$As$_2$ (C) & ~20~ & ~1.017~ & ~5~ & ~200~ \\
    \hline
  \end{tabular}
 \end{center}
\end{table}

A 950~MHz loop-gap resonator, described in detail elsewhere,~\cite{Hardy93-3999} has been used to measure the temperature dependence of the surface reactance $\Delta X_\mathrm{S}$($T$), thus allowing for a determination of $\Delta \lambda(T)$~=~$\lambda(T)$~-~$\lambda$(0), where $\lambda(T)$ is the magnetic penetration depth. Samples are mounted on the end of a temperature-controlled sapphire plate with a small a mount of silicon grease. Microwave magnetic fields are applied parallel to the $ab$-plane, a geometry in which the applied magnetic field at the surface of the sample is almost everywhere equal to the applied field, and the cavity resonance frequency is measured as a function of the sample temperature. In this geometry, screening currents flow around the crystal in both the $\hat a$- and $\hat c$-axis directions. To limit the effects of $\hat c$-axis contamination, platelets with large $a/c$ aspect ratios are preferred. When the system is run as an oscillator (rather than as a conventional resonator), an absolute frequency stability of $\sim$0.1~Hz/min provides sub-angstrom resolution in $\Delta\lambda(T)$ measurements. The samples were cooled to 1.2~K via a $^4$He pumped cryostat, and to 0.4~K via a $^3$He pot and a charcoal sorption pump at 4.2~K. The resonator remained at 1.2~K, in direct contact with a pumped $^4$He bath, throughout the experiment.

Surface resistance $R_\mathrm{S}(\omega,T)$ measurements were made using a recently developed non-resonant broadband microwave apparatus operating between 0.5 and 20~GHz, described in detail elsewhere.~\cite{Turner04-124} As with the above technique, microwave magnetic fields are applied parallel to the $ab$-plane. The microwave power is modulated at low frequency and the measured temperature oscillation of the superconducting sample gives a direct measure of the absorbed power. A reference alloy of Ag:Au, placed in an electromagnetically equivalent position as the superconducting sample, is then used to calibrate the absolute surface resistance of the FeAs crystal.

The temperature dependence of $\Delta R_\mathrm{S}(T)$ was measured at 13~GHz in the axial microwave magnetic fields of the TE$_\mathrm{011}$ modes of a right-circular cylindrical cavity, described elsewhere.~\cite{Hosseini99-1349}

\section{Penetration Depth Results}

\begin{figure}[htpb]
\begin{center}
\includegraphics[bb= 0 15 520 525, width=\columnwidth]{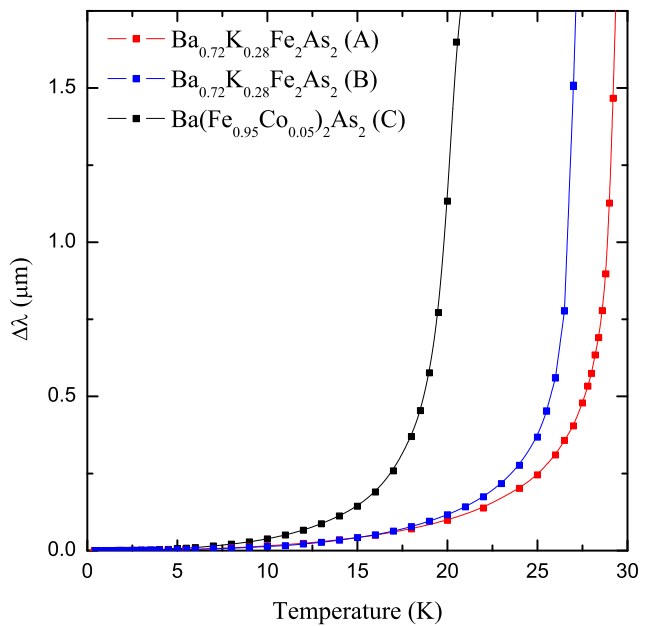}
\caption{(Color online) The change in the London penetration depth, $\lambda(T)$-$\lambda$(0), as a function of temperature for all three samples. Solid lines are a guide to the eye.} \label{fig:LondonPenDep}
\end{center}
\end{figure}

The measured magnetic penetration depth for all three samples is shown in Fig.~\ref{fig:LondonPenDep} as a function of temperature. The measured background frequency shift from an empty sapphire plate with a small amount of silicone grease (approximately equal to that used to hold the sample in place) corresponds to about 0.2~\AA~in $\Delta\lambda$ at base temperature and shows no systematic temperature dependence. The penetration depth of both K-doped crystals (samples A and B) has also been measured in a 12~kHz ac susceptometer, described in detail elsewhere~\cite{Bidinosti00-125014} and, to within calibration uncertainties, agrees with the 950~MHz cavity perturbation measurements.

\begin{figure}[htpb]
\begin{center}
\includegraphics[bb= 7 20 508 521, width=\columnwidth]{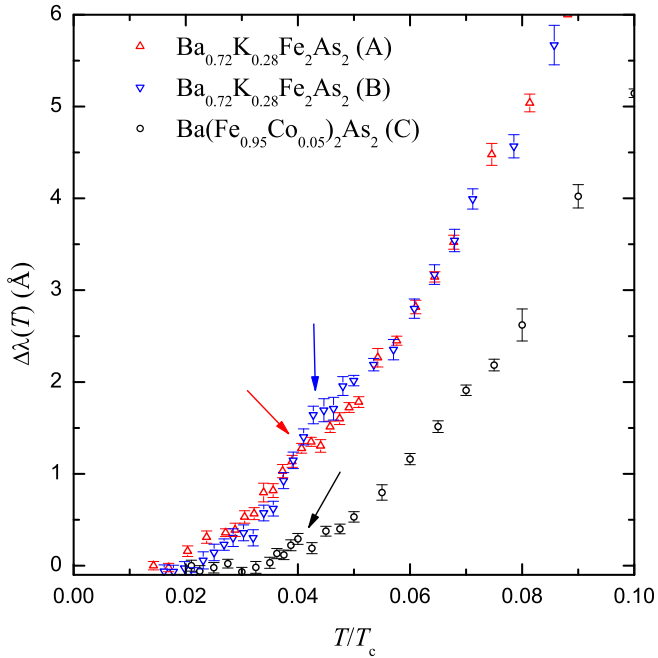}
\caption{(Color online) The anomalous \textquotedblleft bump" in the London penetration depth. For all three samples, the bump occurs at $T/T_\mathrm{c}$~$\approx$~0.04} \label{fig:LowTbumps}
\end{center}
\end{figure}

Before fitting models to the data, an anomaly that appears near $T/T_\mathrm{c}$~$\approx$~0.04 in the measured $\Delta\lambda$ of all three samples is shown in Fig.~\ref{fig:LowTbumps}. That the anomaly scales with $T_\mathrm{c}$ is likely an indication that the feature is relevant and may be related to the onset of magnetic order in the crystals; for example, samples A and B are in the doping range where coexistence of a spin-density wave and superconductivity has been reported,~\cite{Chen09-17006} and sample C has a doping level which is very close to where antiferromagnetism has been observed to coexist with superconductivity.~\cite{Ni08-214515,Chu09-014506} Concern that this feature might be due to a trace contaminant, such as a small particle of superconducting aluminum metal, led us to re-measure after carefully re-cleaning the samples and sapphire. The anomaly persisted, and its origin in the bulk properties of the sample remains uncertain. Due to its small magnitude one can seek to model the overall low temperature behavior without explicitly addressing its presence.

\begin{figure}[htpb]
\begin{center}
\includegraphics[bb= 0 20 510 525, width=\columnwidth]{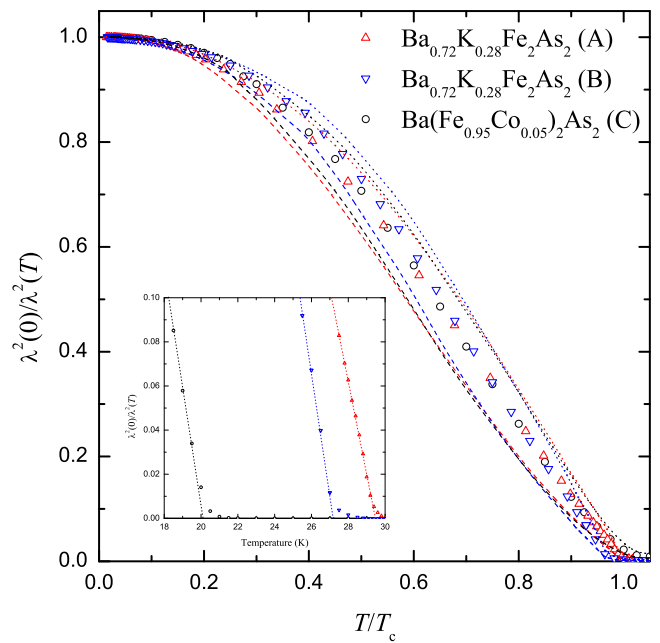}
\caption{(Color online) Extracted superfluid density of all three samples, using three different choices of $\lambda(0)$ (symbols are 2000~{\AA}, dashed lines are 1500~{\AA}, dotted lines are 2500~{\AA}). Inset: the superfluid transition of all three samples. Dotted lines are linear guides to the eye and show that deviations from linearity occur only very near $T_\mathrm{c}$ in the form of \textquotedblleft feet" that indicate minor inhomogeneities in the $T_\mathrm{c}$ of the sample.}\label{fig:SFphasestiffnesswithfeet}
\end{center}
\end{figure}

The temperature dependence of the penetration depth provides access to the superfluid density or, more correctly, the superfluid phase stiffness, provided one has a measure of $\lambda(T_0$), where $T_0$ is some suitably low temperature. Fig.~\ref{fig:SFphasestiffnesswithfeet} shows the extracted superfluid phase stiffness for all three samples.  For the analysis that follows, we take $\lambda(0)$~=~2000~{\AA},~\cite{Li08-107004} but Fig.~\ref{fig:SFphasestiffnesswithfeet} shows that the choice of $\lambda(0)~\pm$~25{\%} does not affect the qualitative features of the superfluid density.

At higher temperatures, the superfluid phase stiffness approaches $T_\mathrm{c}$ approximately linearly, as shown in the inset of Fig.~\ref{fig:SFphasestiffnesswithfeet}, consistent with mean field behavior. Small deviations from linearity occur about 0.5~K from $T_\mathrm{c}$. The onset of curvature is likely a result of small sample inhomogeneity. There appears to be no evidence for the 3D XY critical fluctuation behavior seen in very high quality YBa$_2$Cu$_3$O$_{7-\delta}$ samples~\cite{Kamal98-R8933} where $\lambda^2(0)$/$\lambda^2(T)$ approaches $T_\mathrm{c}$ with infinite, rather than finite, slope. The present iron arsenide crystals under study appear quite homogeneous, especially when compared to the broad \textquotedblleft feet" seen in the superfluid density near $T_\mathrm{c}$ in some measurements.~\cite{Hashimoto09-207001,Prozorov09-582,Martin09-020501}

The temperature dependence upon approaching $T$~=~0 provides the strongest constraints on possible nodes in the superconducting gap, since low temperatures query the lowest energy excitations in the system. At the lowest temperatures, there exists a number of different models, each of which are motivated by physically realizable scenarios. The simple $s$-wave BCS model~\cite{Muhlschlegel59-313} that, at low temperatures, takes the form

\begin{equation}
\frac{\lambda(T)}{\lambda(0)}~\approx~\sqrt{\frac{2 \pi \Delta_{0}}{k_\mathrm{B} T}} \exp{ \left(\frac{-\Delta_{0}}{k_\mathrm{B} T}\right)},\label{eq:swaveBCS}
\end{equation}

\noindent does not fit the data well at all, and additionally yields uncharacteristically small gaps of 2$\Delta$/k$_\mathrm{B}$$T_\mathrm{c}$~$\approx$~1.5 under the best fit conditions for the K-doped samples A and B, and 2$\Delta$/k$_\mathrm{B}$$T_\mathrm{c}$~$\approx$~2.0 under the best fit conditions for the Co-doped sample C. The best fit curves are not shown.

Also worth considering is whether the deviation $\Delta\lambda(T)$ is quadratic in temperature. A $T^2$ temperature dependence of the penetration depth for $T~\ll~T_\mathrm{c}$ is expected for a superconducting gap with line nodes; e.g., $d$-wave symmetry, in the presence of strong scattering, which will create an additional quasiparticle density of states.~\cite{Hirschfeld93-4219} This deviation from otherwise linear behavior of $\Delta\lambda(T)$ predicted for a $d$-wave gap symmetry is expected to occur only below a characteristic temperature $T^\star$, controlled by the impurity concentration. The $T^2$ model we apply to the superfluid phase stiffness takes the form

\begin{equation}
\frac{\lambda^2(0)}{\lambda^2(T)}~\approx~1-\left(\frac{T}{T^\star}\right)^2.\label{eq:Tsquared}
\end{equation}

The quadratic temperature dependence fits to the data work moderately well (not shown) and return values of $T^\star~\approx~T_\mathrm{c}$, which is not surprising, as so much scattering should suppress $T_\mathrm{c}$. It is worth noting that $T^2$ variation of the superfluid density at the lowest temperatures has been observed before, in YBCO films~\cite{Lee93-2419} and crystals~\cite{Bonn93-11314} of relative poor quality. Also worth noting is that within the $s_{\pm}$-wave model with impurity scattering,~\cite{Bang09-054529} some low energy quasiparticle excitations can occur leading to a power law-like dependence, such as a density of states proportional to $E^2$.

\begin{figure}[htpb]
\begin{center}
\includegraphics[bb= 30 25 565 520, width=\columnwidth]{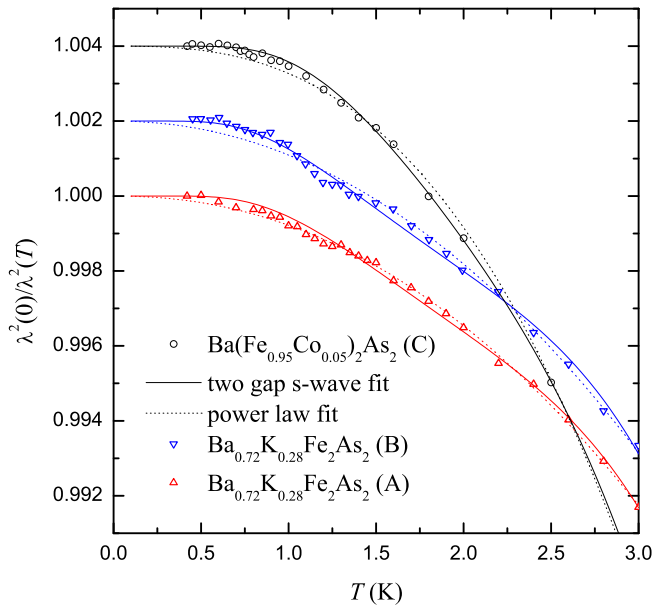}
\caption{(Color online) Two gap $s$-wave (solid line, Eq.~\ref{eq:twogap}) and power law (dashed line, Eq.~\ref{eq:powerlaw}) fits over 0~-~3~K to the superfluid phase stiffness. Notice the very fine scale of $\lambda^2$(0)/$\lambda^2$($T$). For clarity, the superfluid phase stiffness of samples C and B have been shifted vertically by 0.004 and 0.002, respectively.}\label{fig:ModelFits}
\end{center}
\end{figure}

The data can be reasonably well described by a two-gap $s$-wave model,~\cite{Kim02-064511} which might be an appropriate starting point for a superconductor with multiple Fermi surface sheets. At very low temperatures, the change of superfluid phase stiffness with temperature is dominated by the smallest gap, with the larger gap becoming more apparent at high temperatures. The main role of the large gap in modeling the low temperature data is to limit the rapid rise in penetration depth if a small gap were allowed to dominate a large fraction of the Fermi surface in the Brillouin zone. The model applied takes the form

\begin{align}
\frac{\lambda^2(0)}{\lambda^2(T)}~\approx~1 &- x~\sqrt{\frac{2 \pi \Delta_{0,\mathrm{S}}}{k_\mathrm{B} T}} \exp \left(\frac{-\Delta_{0,\mathrm{S}}}{k_\mathrm{B}  T}\right) \notag\\
&- (1-x)~\sqrt{\frac{2 \pi \Delta_{0,\mathrm{L}}}{k_\mathrm{B}  T}} \exp \left(\frac{-\Delta_{0,\mathrm{L}}}{k_\mathrm{B}  T}\right),
\label{eq:twogap}
\end{align}

\noindent where $x~=~\lambda^2(0)/\lambda^2_\mathrm{S}(0)$ and is the fractional contribution of the small gap $\Delta_{0,\mathrm{S}}$ to the total superfluid density. These fits are shown as the solid black lines in Fig.~\ref{fig:ModelFits}. The gap sizes obtained from the two-gap model are small: $\Delta_{0,\mathrm{S}}$ is nearly five times smaller than $\Delta_{0,\mathrm{L}}$, and both are less than the weak-coupling BCS expectation of $2\Delta/k_\mathrm{B}T_\mathrm{c}$~$\approx~3.5$.

Finally, we attempt fits to a power law,

\begin{equation}
\frac{\lambda^2(0)}{\lambda^2(T)}~\approx~1-\left(\frac{T}{T_\mathrm{z}}\right)^n.\label{eq:powerlaw}
\end{equation}

\noindent In clean superconductors, the deviation $\Delta \lambda$ of the penetration depth from its zero-temperature value $\lambda$(0) is proportional to $T^n$ and unveils the topology of the superconducting gap. Specifically, \emph{n}~=~1 for line nodes and \emph{n}~=~2 for point nodes. Non-local effects, impurities, and other defects serve to increase the exponent \emph{n};~\cite{Hirschfeld93-4219} consequently, deciphering the curious exponent \emph{n} relies upon whether it is larger or smaller than 2. The dashed lines of Fig.~\ref{fig:ModelFits} show how a power law with \emph{n}~$\approx$~2.5 can be used to describe the superfluid density of these samples at low temperatures. This power law is in agreement with previous penetration depth measurements.~\cite{Gordon09-100506(R)}

For samples A and B, the data is best described by a power law (as determined by a $\chi^2$ statistic), regardless of the temperature range over which the fit is applied. For sample C, both a power law fit and a two-gap fit describe the data equally well. Table~\ref{tab:2} shows the extracted fit parameters for each model. Note that the fit parameters systematically increase as the fit range is increased for the two-gap model, while there is no such systematic deviation in the parameters for the power law fits. As the 122-system is studied over successively smaller temperature ranges, it seems as though the parameters extracted from two-gap models always seem to tend towards a very small gap on an insignificant fraction of the Fermi surface.

\begin{table}[ht]
 \begin{center}
  \begin{tabular}{|c||c|c|c||c|c|}
  \hline
  Sample A & \multicolumn{3}{|c||}{two gap} & \multicolumn{2}{|c|}{power law}\\
  \hline\hline
  ~ & ~ & ~ & ~ & ~ & ~ \\[-2ex]
  {fit range $T_{\mathrm{max}}$} & $x$ & $\frac{2\Delta_{0,\mathrm{S}}}{k_\mathrm{B} T_\mathrm{c}}$ & $\frac{2\Delta_{0,\mathrm{L}}}{k_\mathrm{B} T_\mathrm{c}}$ & \emph{n} & $T^\star$\\[1ex] \hline
  ~3 K~ & ~0.9\%~ & ~ 0.30~ & ~1.65~ & ~2.15~ & ~28.0 K~ \\
  ~4 K~ & ~1.2\%~ & ~ 0.34~ & ~1.84~ & ~2.22~ & ~25.8 K~ \\
  ~5 K~ & ~1.8\%~ & ~ 0.41~ & ~2.08~ & ~2.25~ & ~24.9 K~ \\
  ~6 K~ & ~2.5\%~ & ~ 0.48~ & ~2.26~ & ~2.31~ & ~23.8 K~ \\
  ~7 K~ & ~3.3\%~ & ~ 0.54~ & ~2.40~ & ~2.35~ & ~23.0 K~ \\
  \hline\hline
  Sample B & \multicolumn{3}{|c||}{two gap} & \multicolumn{2}{|c|}{power law}\\
  \hline\hline
  ~ & ~ & ~ & ~ & ~ & ~ \\[-2ex]
  {fit range $T_{\mathrm{max}}$} & $x$ & $\frac{2\Delta_{0,\mathrm{S}}}{k_\mathrm{B} T_\mathrm{c}}$ & $\frac{2\Delta_{0,\mathrm{L}}}{k_\mathrm{B} T_\mathrm{c}}$ & \emph{n} & $T^\star$\\[1ex] \hline
  ~3 K~ & ~0.8\%~ & ~ 0.29~ & ~1.69~ & ~2.07~ & ~29.3 K~ \\
  ~4 K~ & ~1.2\%~ & ~ 0.34~ & ~1.94~ & ~2.11~ & ~27.9 K~ \\
  ~5 K~ & ~1.8\%~ & ~ 0.42~ & ~2.24~ & ~2.11~ & ~28.1 K~ \\
  ~6 K~ & ~2.6\%~ & ~ 0.50~ & ~2.49~ & ~2.09~ & ~28.5 K~ \\
  ~7 K~ & ~3.3\%~ & ~ 0.56~ & ~2.67~ & ~2.11~ & ~27.9 K~ \\
  \hline\hline
  Sample C & \multicolumn{3}{|c||}{two gap} & \multicolumn{2}{|c|}{power law}\\
  \hline\hline
  ~ & ~ & ~ & ~ & ~ & ~ \\[-2ex]
  {fit range $T_{\mathrm{max}}$} & $x$ & $\frac{2\Delta_{0,\mathrm{S}}}{k_\mathrm{B} T_\mathrm{c}}$ & $\frac{2\Delta_{0,\mathrm{L}}}{k_\mathrm{B} T_\mathrm{c}}$ & \emph{n} & $T^\star$\\[1ex] \hline
  ~3 K~ & ~2.1\%~ & ~ 0.57~ & ~2.24~ & ~2.73~ & ~14.1 K~ \\
  ~4 K~ & ~3.6\%~ & ~ 0.70~ & ~2.62~ & ~2.60~ & ~15.4 K~ \\
  ~5 K~ & ~4.0\%~ & ~ 0.72~ & ~2.75~ & ~2.53~ & ~16.2 K~ \\
  ~6 K~ & ~3.5\%~ & ~ 0.69~ & ~2.66~ & ~2.75~ & ~14.4 K~ \\
  ~7 K~ & ~4.1\%~ & ~ 0.74~ & ~2.71~ & ~2.70~ & ~14.7 K~ \\
  \hline
  \end{tabular}
  \caption{Parameters from two gap $s$-wave and power law fits. For sample A, the data is best described by the power law $T^n$ with \emph{n}~$\approx$~2.26~$\pm$~0.08. For sample B, the data is best described by the power law $T^n$ with \emph{n}~$\approx$~2.10~$\pm$~0.02. For sample C, both a two-gap $s$-wave model and a power law model $T^n$ with \emph{n}~$\approx$~2.66~$\pm$~0.09 describe the data equally well over all temperature ranges.}\label{tab:2}
 \end{center}
\end{table}

\section{Microwave Spectroscopy Results}

Well below $T_\mathrm{c}$, most superconducting properties should depend strongly on the quasiparticle excitations near the nodes (or minima) of the gap function. Important signatures of nodal-quasiparticle transport emerge in the frequency dependence of the in-plane microwave conductivity, $\sigma_1$($\omega$,$T$), which is extracted from measurements of $R_\mathrm{S}(\omega,T)$. Specifically, at low temperature, the surface resistance is related to the conductivity via the relation:~\cite{Chang89-4299}

\begin{equation}
R_S(\omega, T) \simeq \frac{1}{2} \mu_0^2 \omega^2 \lambda^3(T) \sigma_1(\omega,T).\label{eq:Rs}
\end{equation}

\begin{figure}[htpb]
\begin{center}
\includegraphics[bb= 70 35 680 530, width=\columnwidth]{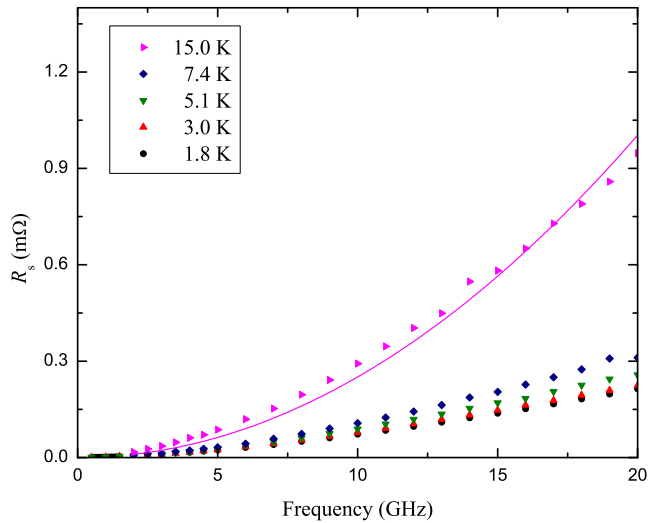}
\caption{(Color online) In-plane surface resistance of sample B from 0.5 to 20~GHz. The loss of the sample B at 20~GHz and base temperature is a factor of 3.5 less than that of sample A. The solid line is $\propto\omega^2$.}\label{fig:RsB}
\end{center}
\end{figure}

\begin{figure}[htpb]
\begin{center}
\includegraphics[bb= 70 35 680 530, width=\columnwidth]{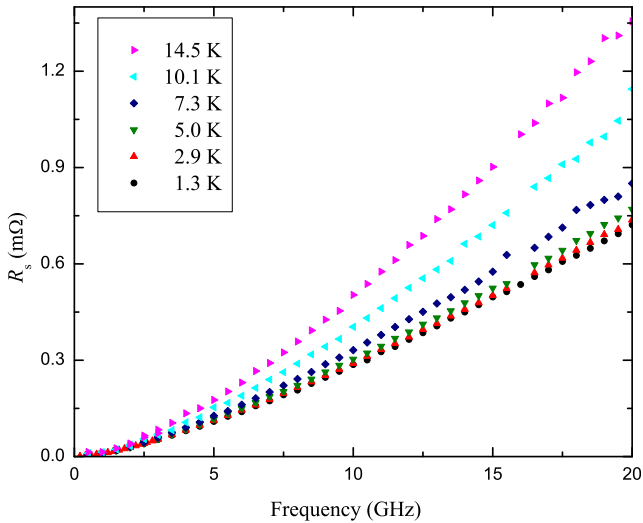}
\caption{(Color online) In-plane surface resistance of sample A from 0.5 to 20~GHz.}\label{fig:RsA}
\end{center}
\end{figure}

$R_\mathrm{S}(\omega,T)$ spectra of the two \FeAs~samples B and A are shown in Figs.~\ref{fig:RsB} and~\ref{fig:RsA}, respectively. There is a significant difference in the low temperature surface resistance between the two samples. At 20~GHz and 3~K, the $R_\mathrm{S}$ of the thick sample A is roughly 3.5 times larger than that of the thin sample B. The same is not true, however, at higher temperatures. At 20~GHz and 15~K, the $R_\mathrm{S}$ of the two samples are now much closer and differ only by 30\%. This could be the result of an extrinsic temperature-independent residual loss (of unknown origin), dominant when the sample itself shows little loss and obscured when the sample loss is sufficiently high.

\begin{figure}[htpb]
\begin{center}
\includegraphics[bb= 70 20 655 540, width=\columnwidth]{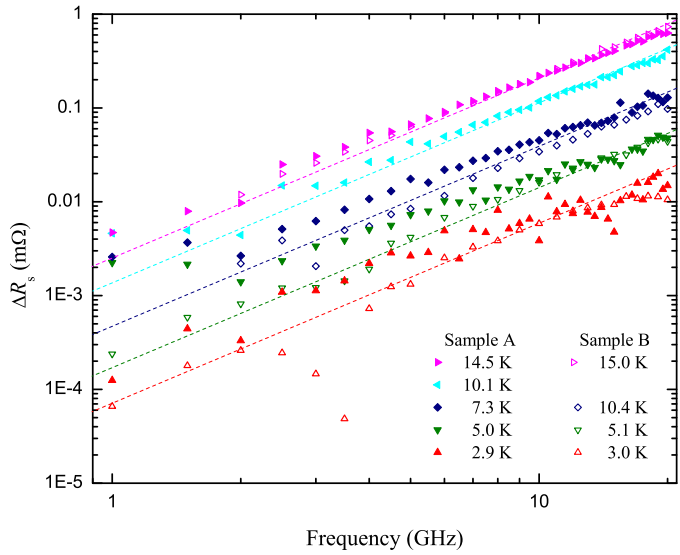}
\caption{(Color online) $\Delta R_\mathrm{S}(\omega,T)=R_\mathrm{S}(\omega,T)-R_\mathrm{S}(\omega,T_0)$ for samples A (solid symbols) and B (open symbols). $\Delta R_\mathrm{S}(\omega,T)$ is seen to be the same for both samples indicating that sample A (the thicker of the two) has an extra temperature independent loss. The dashed lines have a slope of 2 ($\Delta R_\mathrm{S}\propto\omega^2)$.}\label{fig:DeltaRs}
\end{center}
\end{figure}

Another indication that the low temperature loss might be extrinsic is that at $T$~=~15~K in the lower loss sample, shown in Fig.~\ref{fig:RsB}, $R_\mathrm{S}$ is proportional to $\omega^2$, a result expected if there is sufficient scattering to yield conductivity $\sigma_1(\omega)$ that is independent of frequency in the microwave range. Elimination of the residual component is performed by subtracting the loss at the lowest measured temperature, a procedure which has been shown to work for $s$-wave superconductors.~\cite{Turneaure} Doing this reveals that the absolute temperature evolution of $R_\mathrm{S}(\omega,T)$ is the same for both samples, as highlighted by the plot of $\Delta R_\mathrm{S}(\omega,T)$~=~$R_\mathrm{S}(\omega, T)~-~R_\mathrm{S}(\omega, T_0)$ in Fig.~\ref{fig:DeltaRs}, supporting the hypothesis of an extrinsic residual loss. Specifically, Fig.~\ref{fig:DeltaRs} demonstrates that the resulting loss after subtraction seems to be intrinsic, in that it has the same temperature and $\omega^2$ dependence in both samples. Making this subtraction, however, removes the possibility of being able to state whether or not some intrinsic residual resistivity actually does exist.

In the \FeAs~samples A and B, it is very likely that the K-dopants are the dominant source of quasiparticle scattering. With such a high concentration (28\%) of dopants, it is reasonable to expect that the widths of the quasiparticle conductivity spectra lie well outside of our measurement bandwidth. In this case, the in-plane microwave conductivity $\sigma_1(\omega,T)\approx 2R_\mathrm{S}(\omega,T)/\mu_0^2\omega^2\lambda^3(T)$ would appear frequency independent, resulting in the $R_\mathrm{S}(\omega,T)\sim\omega^2$ as shown by the lines in Fig.~\ref{fig:DeltaRs}.

There exist a number of factors that might account for the observed excess loss in the \FeAs~samples, and surface contamination (perhaps a barium oxide layer) seems to be the most likely source. Both samples A and B were cleaved from a larger crystal before they were measured, with one potentially significant difference: while one surface of sample A was cleaved just prior to the start of measurements, the opposite surface had been cleaved weeks earlier allowing for the possibility of enhanced degradation of that surface. Sample B, on the other hand, had both surfaces freshly cleaved immediately before measurements began.

\begin{figure}[htpb]
\begin{center}
\includegraphics[bb= 0 0 145 145, width=\columnwidth]{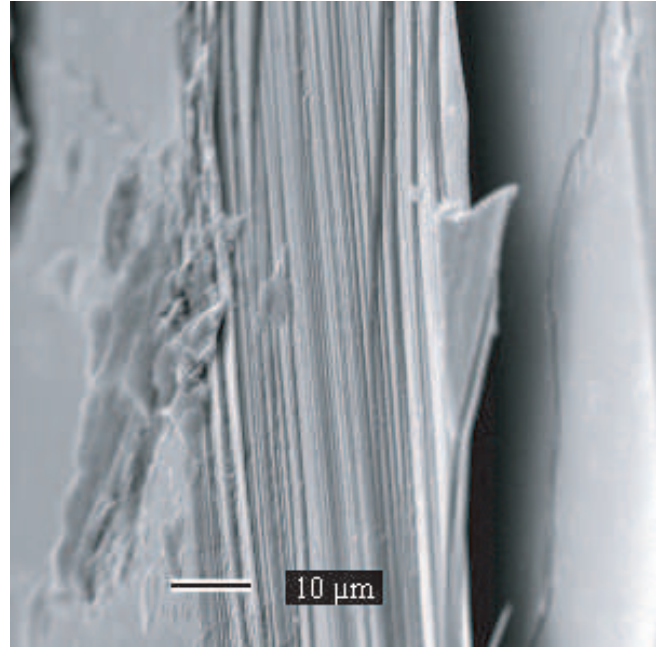}
\caption{Scanning electron microscope image of the edge of a platelet \FeAs crystal, showing delaminated edges.}\label{fig:FeAsSEM}
\end{center}
\end{figure}

A second possible source of the observed excess loss is the presence of delaminated edges on the samples. Sample edges that resemble the unbound leafy edge of a book present obvious problems for microwave measurements. In the extreme case, the microwave fields would be allowed to penetrate the regions between loosely connected \textquotedblleft sheets" of the sample. Currents that loop around the sample are presented with a tortuous path near the edges that increases the effective surface area of the sample. In fact, preliminary measurements of $\Delta\lambda(T)$ by our group on early samples with obvious edge problems showed a linear temperature dependence that had an unrealistic slope of ~$\sim$~200~\AA/K, cf. YBCO that has $\Delta\lambda(T)~\approx~$4~\AA/K.~\cite{Kamal98-R8933} Scanning electron microscope images, as in Fig~\ref{fig:FeAsSEM}, have confirmed that some crystals do show signs of delaminated edges with sheets~$\approx~1~-~2~\mu$m thick. This effect is therefore suppressed for the thinner \FeAs~sample B (and the \FeCoAs~sample C), compared to the thicker sample A, measured in this work. This effect touches upon a slightly different issue than the surface degradation problem mentioned above, since it could affect the data in a multiplicative way, not just additively; however, it is an effect worth noting as yet another reason to work on only very thin and freshly cleaved samples.

It also seems possible that intrinsic $\hat c$-axis contamination was providing the excess loss. However, the thicker sample A was cut into two pieces such that any $\hat c$-axis contribution to the $R_\mathrm{S}(T)$ measurements would be enhanced (doubled) but such enhancement was not observed in the measurements.

Despite falling well outside the measurement bandwidth, it is possible to obtain an estimate of the spectral width $\Gamma$ through the oscillator-strength sum rule. (Our broadband system measures $\sigma_1(\omega,T)$ out to $\omega$/2$\pi$~=~20~GHz and in this spectral region we find $\sigma_1(\omega,T)$ is independent of $\omega$ and therefore no direct measure of the width of the conductivity spectrum can be obtained.) The oscillator-strength sum rule is given by:

\begin{equation}
\frac{n_{\mathrm{n}}(T)e^2}{m^*}~=~\frac{2}{\pi}\int_0^\infty\sigma_1(\omega,T)d\omega.\label{eq:sumrule}
\end{equation}

\begin{figure}[htpb]
\begin{center}
\includegraphics[bb= 80 260 670 530, width=\columnwidth]{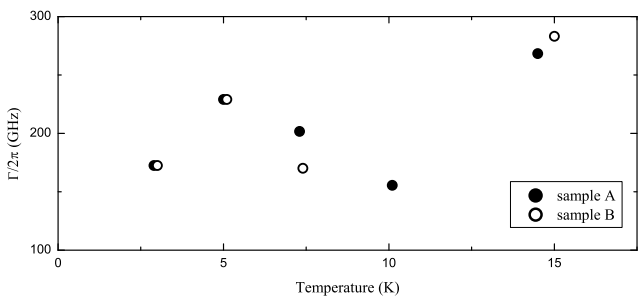}
\caption{Estimates of the quasiparticle conductivity spectral widths $\Gamma/{2 \pi}$ of samples A and B as a function of temperature.}\label{fig:spectralwidth}
\end{center}
\end{figure}

\noindent As the temperature is raised, any spectral weight depleted from the superfluid density $n_\mathrm{s}$($T$) must reappear as an increase in the frequency-integrated quasiparticle conductivity. One can approximate the integral on the right-hand side of Eq.~\ref{eq:sumrule} to be equal to $\sigma_0\Gamma$, and the left-hand side of Eq.~\ref{eq:sumrule} is determined by partitioning the conduction electron density $n$ into a superfluid density $n_\mathrm{s}$ and a normal-fluid density $n_\mathrm{n}$~=~$n$~-~$n_\mathrm{s}$~$\sim$~1/$\lambda^2(0)-1/\lambda^2(T)$, corresponding to quasiparticles thermally excited from the condensate. In so doing, we make use of $\sigma_0\approx2R_\mathrm{S}(\omega,T)/\mu^2_0\omega^2\lambda^3(T)$ being the extrapolated value of $\sigma_1$($\omega~\rightarrow$~0) and $\Gamma$ as a measure of the spectral width of $\sigma_1(\omega,T)$. From this generalized two-fluid model,~\cite{Berlinsky93-4074} the value of $n_\mathrm{s}$ (and its temperature dependence) is determined from measurements of the magnetic penetration depth and recalling the definition of the London penetration depth $\lambda_\mathrm{L}^{-2}$=$\mu_0 n_\mathrm{s} e^2/m^*$. The temperature dependence of the spectral widths of samples A and B so determined are shown in Fig.~\ref{fig:spectralwidth}. These values of $\Gamma/2\pi$ are more than an order of magnitude larger then what has been observed in YBCO~\cite{Turner03-237005} and show no clear temperature dependence; their magnitudes, however, are comparable to the low temperature scattering rates of Bi$_2$Sr$_2$CaCu$_2$O$_{8+\delta}$ derived from THz spectroscopy.~\cite{PhysProp}

\begin{figure}[htpb]
\begin{center}
\includegraphics[bb= 65 5 670 550, width=\columnwidth]{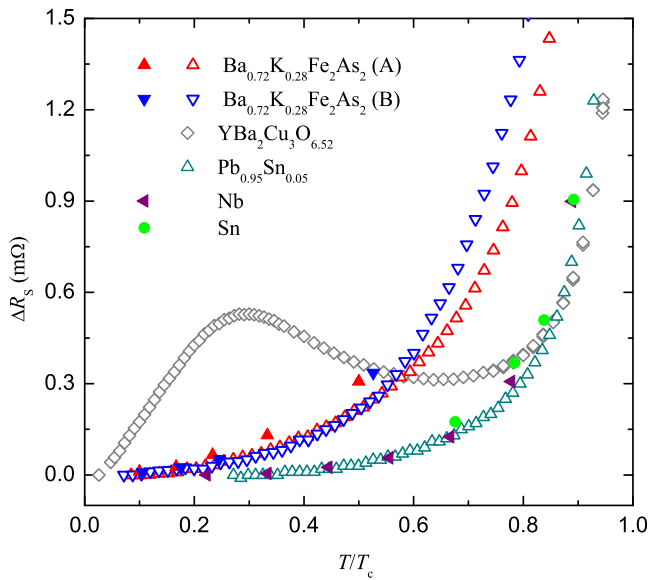}
\caption{(Color online) $\Delta R_\mathrm{S}(\omega,T)$ at 13~GHz as a function of $T/T_\mathrm{c}$ for several superconductors. Solid symbols were taken using the broadband spectrometer. Open symbols were taken using a 13~GHz resonator. The conventional $s$-wave superconductors (Pb$_{0.95}$Sn$_{0.05}$, Nb, and Sn) exhibit a very weak temperature dependence for $T/T_\mathrm{c}<0.4$. The \FeAs~samples obey a power law $\Delta R_\mathrm{S}(\omega,T)\propto T^\gamma$ down to the lowest measurement temperature.}\label{fig:13GHz}
\end{center}
\end{figure}

Finally, Fig.~\ref{fig:13GHz} shows the temperature dependence of $\Delta R_\mathrm{S}(\omega,T)$ at 13~GHz for a variety of superconductors. For each of the conventional BCS $s$-wave superconductors (Pb$_{0.95}$Sn$_{0.05}$, Nb, and Sn), $\Delta R_\mathrm{S}(\omega,T)$ scales as $T/T_\mathrm{c}$ and shows a very weak temperature dependence below $T/T_\mathrm{c}=0.4$. The \FeAs~data, on the other hand, shows no signs of flattening and obeys a power law $R_\mathrm{S}(\omega,T)\propto T^n$ down to the lowest measurement temperature. Fits up to $T/T_\mathrm{c}=0.5$ give $n~\approx~2.5~\pm~0.3$, which is the temperature dependence expected for a normal fluid density that tracks the loss of superfluid density. Rigorous comparisons to the low-$T$ power law fits to the superfluid phase stiffness, however, are not possible because the $\Delta R_\mathrm{S}(\omega,T)$ do not extend to low enough base temperatures and the resolution of the measurement at the lowest temperatures is inadequate. Nevertheless, it is clear that the overall temperature dependence of the surface resistance of the \FeAs~samples differs significantly from that of the $s$-wave BCS superconductors.

The surface resistance of ortho-II ordered YBa$_2$Cu$_3$O$_{6.52}$ (grey open symbols of Fig.~\ref{fig:13GHz}) shows a low-temperature peak (as does the quasiparticle conductivity). This peak results from a competition between the temperature dependencies of the normal fluid density and of the quasiparticle scattering rates; at lower temperatures it is the reduced normal fluid density which wins out over large scattering rates, while the converse is the case at higher temperatures.~\cite{Hosseini99-1349,Harris06-104508} This feature is known to be extremely sensitive to disorder and/or impurities,~\cite{Bonn94-4051} and considering that the spectral width of the quasiparticle conductivity of \FeAs~is estimated to be an order of magnitude larger than it is in ortho-II ordered YBa$_2$Cu$_3$O$_{6.52}$, the absence of a peak in $R_\mathrm{S}(\omega,T)$ is not surprising, even though there may be a modest decrease in the scattering rate below $T_\mathrm{c}$.

Because the broadband measurements have revealed an excess extrinsic loss in our \FeAs~samples, it is not possible to obtain reliable quasiparticle conductivities from our $R_\mathrm{S}(T)$ measurements. The exact behavior of $\sigma_1(T)$ necessitates simultaneous measurements of $R_\mathrm{S}(T)$ and $X_\mathrm{S}(T)$. Limited to what we can extract, an approximated quasiparticle conductivity, we see no evidence of a coherence peak just below $T_\mathrm{c}$. This is in contrast to what has been previously reported,~\cite{Hashimoto09-207001} but those conclusions were drawn from a broad, nearly 5~K wide enhancement in the quasiparticle conductivity just below $T_\mathrm{c}$ in a sample whose superconducting transition temperature width $\Delta T_\mathrm{c}$~$\approx$~2.5~K. The behavior of $\sigma_1(T)$ is known to be highly susceptible to inhomogeneity broadening of the superconducting transition. Similar quasiparticle conductivity enhancements below $T_\mathrm{c}$ were observed early on in the cuprates~\cite{Bonn93-11314,Holczer91-152} but were later attributed to broadened superconducting transitions.~\cite{Glass91-4495,Klein92-2407,Olson91-2406} Our samples have $\Delta T_\mathrm{c}$~$\approx$~0.5~K and we extract a much narrower, only 0.5~K wide, sharp cusp that can be attributed to superconducting critical fluctuations, rather than evidence of a coherence peak. We do, however, observe the same lower peak associated with the quasiparticle scattering rate.

\section{Conclusions}

We have measured the microwave electrodynamics of single crystal iron-based superconductors \FeAs~(hole-doped, $T_\mathrm{c}$~$\approx$~30~K) and \FeCoAs~(electron-doped, $T_\mathrm{c}$~$\approx$~20~K), by cavity perturbation and broadband spectroscopy.

Sample quality and homogeneity were confirmed with a SQUID magnetometer via the width of the superconducting transition as a function of applied field. Penetration depth measurements further confirm that these samples were homogeneously doped, especially when compared with the sharpness of the transitions in other published superfluid measurements, which show a wide range of \textquotedblleft feet" above $T_\mathrm{c}$ .

Using a 950~MHz loop-gap resonator (with results checked against a 12~kHz ac susceptometer), we were able to measure $\Delta\lambda(T)$ with sub-angstrom resolution down to 0.4~K. An anomaly in $\Delta\lambda$, common to all three samples, occurred at $T$/$T_\mathrm{c}$~$\approx$~0.04. While the source of this feature is not currently understood, we have ruled out contaminants on the sample surface and systematic background signals associated with the experimental apparatus. Four separate mathematical models were fit to the superfluid phase stiffness, extracted from the $\Delta\lambda(T)$ data, over a temperature range which extends up to 20\% of $T_\mathrm{c}$. For the \FeAs~samples, the behavior is best described by a power law with $T^{2.5}$. The data was also relatively well-described with a two-gap $s$-wave model, although the fit parameters ($x$, $\Delta_{0,\mathrm{S}}$, and $\Delta_{0,\mathrm{L}}$) all systematically increase as the fit range is increased. Moreover, $x$ is very small resulting in unbalanced contributions to 1/$\lambda^2$(0) from the small and large gaps, and the gap values are always smaller than the BCS weak-coupling result: 2$\Delta_{0,\mathrm{L,S}}$/k$_{\mathrm{B}}T_{\mathrm{c}}$~$<$~3.5. These are perhaps all indications that the $s$-wave two-gap scenario is not the correct picture for the 122-pnictides. Alternatively, if interband coupling is strong, a multiple gap scenario could persist, however it may not then be appropriate to treat the gaps independently. For the \FeCoAs~sample, both the power law and the two-gap $s$-wave models seemed to work equally well. None of the data is very well-described with a $T^2$ fit.

We have also taken broadband surface resistance measurements, from 1-20~GHz, which reveal a sample dependent residual loss in the \FeAs~samples whose origin is unclear. An additive extrinsic loss is posited, from which an underlying intrinsic behavior common to both samples can be recovered by subtracting the lowest temperature measure of $R_\mathrm{S}$($\omega$). In so doing, both \FeAs~samples follow a nearly $\omega^2$ trend. The direct implication is that the quasiparticle conductivity is frequency independent up to our highest frequency of 20~GHz. This is consistent with our estimates of the quasiparticle scattering rates which are in the range of 150~-~270~GHz, much higher than those seen in YBCO but comparable to the rates observed in BSSCO.

\section{Acknowledgments}

The work done at the University of British Columbia was funded by the Natural Sciences and Engineering Research Council of Canada and the Canadian Institute for Advanced Research. The work done at the National Laboratory for Superconductivity was supported by the Natural Science Foundation of China, the Ministry of Science and Technology of China (No. 2006CB601000, No. 2006CB921802), and the Chinese Academy of Sciences.

\end{document}